\documentclass[final,1p]{elsarticle}

\usepackage{epsfig}

\usepackage{amssymb}

\usepackage{lineno}

\journal{NIM A}

\begin{document}

\begin{frontmatter}



















\author[label1]{C.~Adloff}
\author[label1]{Y.~Karyotakis}

\author[label2]{J.~Repond}

\author[label3]{J.~Yu}

\author[label4]{G.~Eigen}

\author[label5]{C.~M.Hawkes}  
\author[label5]{Y.~Mikami}
\author[label5]{O.~Miller}
\author[label5]{N.~K.Watson}
\author[label5]{J.A.~Wilson}

% ...

\address[label1]{Laboratoire d'Annecy-le-vieux de Physique des Particules, 
Chemin du Bellevue BP 110, F-74941 Annecy-le-Vieux Cedex, France}

\address[label2]{Argonne National Laboratory, 9700 S.\ Cass Avenue, Argonne, IL 60439-4815, USA}

\address[label3]{University of Texas, Arlington, TX 76019, USA}

\address[label4]{University of Bergen, Inst. of Physics, Allegaten 55, N-5007 Bergen, Norway}

\address[label5]{University of Birmingham, School of Physics and Astronomy, Edgbaston, Birmingham B15 2TT, UK}

\address[label6]{University of Cambridge, Cavendish Laboratory, J J Thomson Avenue, CB3 0HE, UK}

\address[label7]{Universit\'{e} Hassan II A\"{\i}n Chock, Facult\'{e} des sciences. B.P. 5366 Maarif, Casablanca, Morocco}

\address[label8]{Laboratoire de Physique Corpusculaire de Clermont-Ferrand (LPC), 24 avenue des Landais,
63177 Aubi\`ere CEDEX, France}
\cortext[cor1]{Corresponding author, e-mail: carlogan@in2p3.fr}

\address[label9]{NICADD, Northern  Illinois University, Department of Physics, DeKalb, IL 60115, USA}

\address[label10]{Laboratoire de Physique Subatomique et de Cosmologie - Universit\'{e} Joseph Fourier Grenoble 1 - CNRS/IN2P3 - Institut Polytechnique de Grenoble, 53, rue des Martyrs, 38026 Grenoble CEDEX, France}

\address[label11]{DESY, Notkestrasse 85, D-22603 Hamburg, Germany}

\address[label12]{Univ. Hamburg, Physics Department, Institut f\"ur Experimentalphysik, Luruper Chaussee 149,
22761 Hamburg, Germany}

\address[label13]{University of Heidelberg, Fakultat fur Physik und Astronomie, Albert Uberle Str. 3-5 2.OG Ost, D-69120 Heidelberg, Germany}

\address[label14]{University of Iowa, Dept. of Physics and Astronomy, 203 Van Allen Hall, Iowa City, IA 52242-1479, USA}

\address[label15]{Chonbuk National University, Jeonju, 561-756, South Korea}

\address[label16]{Kangnung National University, HEP/PD, Kangnung, South Korea}

\address[label17]{Department of Physics, Kobe University, Kobe, 657-8501, Japan}

\address[label18]{Imperial College, Blackett Laboratory, Department of Physics, Prince Consort Road, London SW7 2BW, UK }

\address[label19]{Department of Physics and Astronomy, University College London, Gower Street, London WC1E 6BT, UK}

\address[label20]{Royal Holloway University of London, Dept. of Physics, Egham, Surrey TW20 0EX, UK}

\address[label21]{Universit\'{e} de Lyon, F-69622, Lyon, France ; Universit\'{e} de Lyon 1, Villeurbanne ;
CNRS/IN2P3, Institut de Physique Nucl\'{e}aire de Lyon}

\address[label22]{The University of Manchester, School of Physics and Astronomy, Schuster Lab, Manchester M13 9PL, UK}

\address[label23]{Institute of Theoretical and Experimental Physics, B. Cheremushkinskaya ul. 25,
RU-117218 Moscow, Russia}

\address[label24]{M.V. Lomonosov Moscow State University , D.V.Skobeltsyn Institute of
Nuclear Physics (SINP MSU), 1/2 Leninskiye Gory, Moscow, 119991, Russia}

\address[label25]{Max Planck Inst. f\"ur Physik, F\"ohringer Ring 6,
D-80805 Munich, Germany}
\fntext[labelnum]{Now at Universit\"at G\"ottingen}

\address[label26]{Laboratoire de L'acc\'elerateur Lin\'eaire, Centre d'Orsay, Universit\'e de Paris-Sud XI,
BP 34, B\^atiment 200, F-91898 Orsay CEDEX, France}

\address[label27]{\'Ecole Polytechnique, Laboratoire Leprince-Ringuet (LLR),
Route de Saclay, F-91128 Palaiseau, CEDEX France}

\address[label28]{Pohang Accelerator Laboratory, Pohang 790-784, South Korea}

\address[label29]{Charles University, Institute of Particle \& Nuclear Physics, V Holesovickach 2,
CZ-18000 Prague 8, Czech Republic}

\address[label30]{Institute of Physics, Academy of Sciences of the Czech Republic, Na Slovance 2,
CZ-18221 Prague 8, Czech Republic}

\address[label32]{Centre National de l'Energie, des Sciences et des Techniques Nucl\'{e}aires, 
B.P. 1382, R.P. 10001, Rabat, Morocco}

\address[label33]{Ewha Womans University, Dept. of Physics, Seoul 120, South Korea}

\address[label34]{Sungkyunkwan University, 300 Cheoncheon-dong, Jangan-gu, Suwon, Gyeonggi-do  440-746, South Korea}

\address[label35]{Yonsei  University, Dept. of Physics, 134 Sinchon-dong, Sudaemoon-gu, Seoul 120-749, South Korea}

\title{Response of the CALICE Si-W Electromagnetic Calorimeter  Physics Prototype to Electrons}





\begin{abstract}
A prototype Silicon-Tungsten electromagnetic calorimeter  (ECAL) for an International Linear Collider (ILC) detector 
was installed and tested during summer and autumn 2006 at CERN. The detector 
had 6480 silicon pads of dimension 1$\times$1 cm$^2$. Data were collected with electron 
beams in the energy range 6 to 45 GeV.  The analysis described in this paper 
focuses on electromagnetic shower reconstruction and characterises the  \mbox{ECAL}
response to electrons in terms of energy resolution and linearity. The detector is linear to within approximately the 1\% level 
and  has a  relative energy resolution of
$(16.6 \pm 0.1)/ \sqrt{ E({\mathrm{GeV}})} \oplus 1.1\pm 0.1$~(\%). The spatial 
uniformity and the time stability of the ECAL  are  also addressed.
\end{abstract}


\end{frontmatter}


\section{Introduction}
\label{introd}

The CALICE Collaboration is conducting R\&D into calorimetric systems for the
ILC~\cite{ILC-RDR} --- 
a proposed e$^+$e$^-$ linear collider intended to operate at a centre of mass 
energy ranging up to the TeV scale.  
The physics scope at the ILC includes  precise measurements  of the triple- and quartic-gauge bosons interactions, as well as the complete characterisation of the Higgs  and top quark sectors.  In addition, hints of physics beyond the Standard Model  could be addressed in a model-independent way. 

The final states are typically multiple hadronic jets, accompanied frequently by low momentum leptons and/or missing energy. In such cases, lepton identification is difficult and the signature of the final states of interest  relies  on the identification of Z or/and W bosons in their decay modes into two jets. 
In order to distinguish them efficiently, a jet energy resolution close to $30\%/\sqrt{E/\mathrm{GeV}}$ has to be achieved~\cite{ILC-RDR}. A precise reconstruction of the jet direction is also required.  These are the main requirements driving  the detector design  in general at the ILC and the calorimetry design in particular.

The target jet energy resolution  represents an improvement by a factor of two over the best
obtained in previous detectors. Moreover the detection environment becomes more complex with increasing centre of mass energy.  A promising way to achieve this increase in resolution is through 
designing a detector system optimised for 
the so called ``{\it particle flow}'' approach~\cite{pflow}, which relies on the separate
reconstruction of as many particles in the jet as possible, 
using the most suitable detector systems. 

The success of such an algorithm depends on the quality of the pattern recognition 
in the calorimeters.  For particle flow, a high spatial granularity is therefore
as important as the intrinsic energy resolution for single particles.
Furthermore, the overall design of the detector (tracking, electromagnetic 
and hadronic calorimetry) needs to be considered in a coherent way.

The  design of the ILC detectors can be optimised using  Monte Carlo 
simulations, but in order to do this, it is crucial to validate the Monte 
Carlo tools with data.  Therefore, the R\&D of the CALICE Collaboration 
has two broad aims.
The first is to construct realistic calorimeter prototypes, and learn about their 
operation and behaviour in beam tests.
The second objective is to compare the data with 
Monte Carlo simulations using the same tools
used for the full detector. This is especially important in the case of hadronic
showers, where many models are available, 
which make differing predictions for the calorimeter response.  
The CALICE plan is to expose complete calorimeter systems (electromagnetic and hadronic,
using various technologies) to test beams of 
electrons, muons and hadrons. To this end, a first round of beam tests was
performed at DESY and CERN in summer 2006, using a Silicon-Tungsten sampling
electromagnetic calorimeter~\cite{HardPaper}, followed by a hadron calorimeter composed of 
iron and scintillator tiles~\cite{TB:HCAL}, 
and then a Tail Catcher and Muon Counter (TCMT) of iron instrumented with scintillator strips~\cite{TB:TCMT}. 

In this paper, we report results of exposure of the 
prototype  to electron beams in the energy range 6-45 GeV at the CERN H6 beam line~\cite{TB:CERN}.  In 
Section~\ref{sec:TB} we outline the layout of the beam tests.  The  ECAL is briefly described  in Section~\ref{sec:ECALHard} and  some key technical aspects of its performance are highlighted. Section~\ref{sec:MC} summarises the  Monte Carlo simulation. Features of the electron beam data are reviewed in Section~\ref{sec:Selection} and  the uniformity across the detector is addressed. The results of the energy measurement together with some of their systematic uncertainties are presented in Section~\ref{sec:Response} for the detector areas of uniform response.

Several other studies of the prototype are ongoing, exploiting its unprecedentedly fine segmentation and capacity to observe shower development in detail.  These studies will be reported in subsequent publications.


\section{Experimental Setup}
\label{sec:TB}
\begin{figure}[htbp]
  \centering
  \mbox{
    \epsfig{file=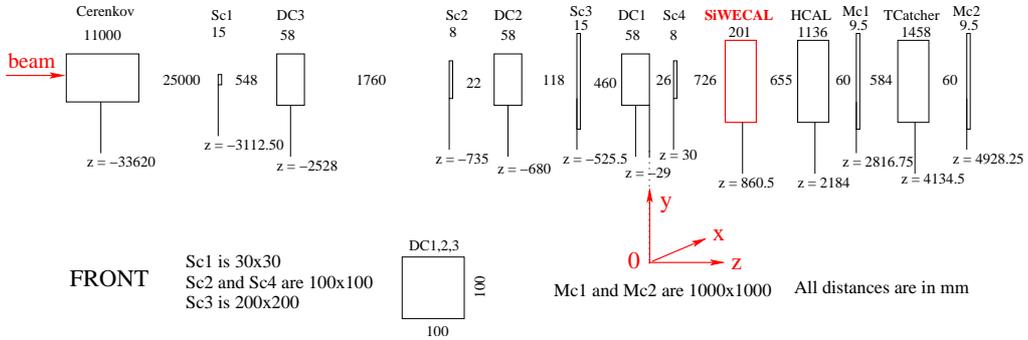, width=\linewidth}
  }
  \caption{Sketch of the CERN test beam setup. The the right handed coordinate system used hereafter is indicated.}
  \label{fig:TB:setup}
\end{figure}

A sketch of the CERN  H6~\cite{TB:CERN} test beam setup is presented in Figure~\ref{fig:TB:setup} and
a detailed description of the detectors can be found in~\cite{HardPaper}. 
The coordinate system used is right handed. The surface of the drift chamber (DC1) closest to the ECAL  defines the origin, the $z$ axis is the beam axis and  $x$ and $y$ the horizontal and upward-vertical, respectively.

The physics program and the overall  electron, pion and muon statistics collected are extensively discussed in~\cite{HardPaper}. 
The   beam trigger was defined by the coincidence signal of two scintillator counters.  In addition, 
three drift chambers were used to monitor the beam. A threshold \v{C}erenkov detector was also available for $e/\pi$ discrimination. 

\begin{figure}[htbp]
  \centering
  \mbox{
    \epsfig{file=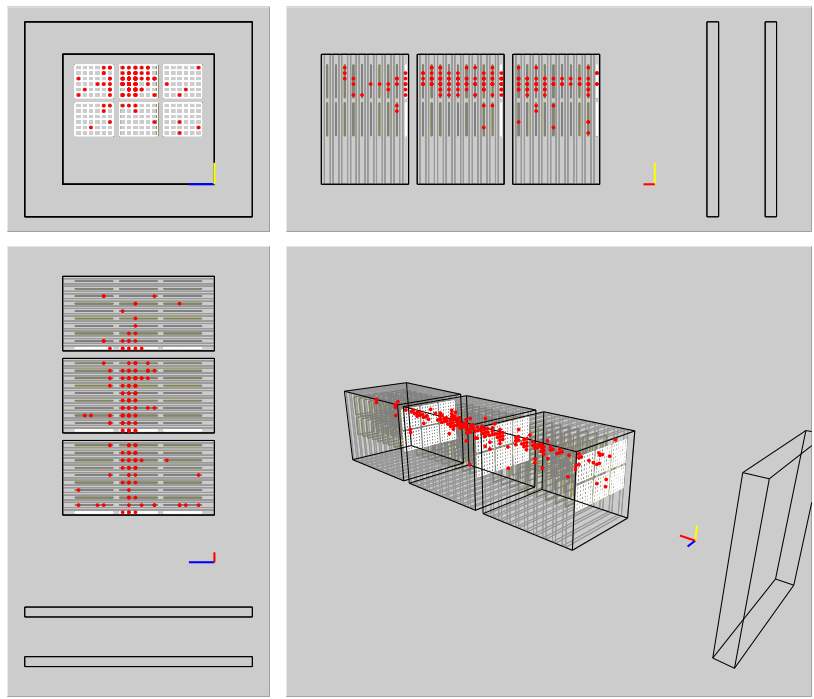, width=0.8\linewidth}
  }
  \caption{ A typical 10~GeV electron shower in the ECAL. The  displayed detector cells have energies higher than  
0.5~MIPs. Layout not to scale.}
  \label{fig:eventdisplay}
\end{figure}

This paper presents the measurement of the ECAL response to electrons normally incident on the calorimeter surface.
The  event 
display for one of these events is shown in Figure~\ref{fig:eventdisplay}, where the energy of the  hits is measured in MIP units, one MIP being
the signal left by a minimum ionising particle.

\section{The Si-W ECAL prototype}
\label{sec:ECALHard}

A detailed description of the ECAL hardware is given in~\cite{HardPaper}, along with details of the commissioning and 
a number of technical features of the system calibration and performance.
The ECAL prototype consisted of 30 layers of tungsten, the first ten of 
thickness 1.4~mm, the next ten of 2.8~mm and the last ten of 4.2~mm, comprising 
24 radiation lengths in total at normal incidence.  The development of the 
showers was sampled using 30 layers of silicon PIN diode pads 
interleaved between the tungsten plates.  The silicon 
thickness was 525~$\mu$m, with each pad having a transverse area
of 1$\times$1~cm$^2$. The sensors were implemented on 4 square inch wafers, segmented into 6$\times$6 pads.
At the time of the 2006 CERN beam tests, each layer consisted of a 
3$\times$2 array of wafers, {\it i.e.}\ 18 pads horizontally and 12 pads vertically, leading
to a total of  6480 pads for the ECAL.

Blocks of random triggers were recorded during data taking in order 
to monitor pedestals and noise. Short term changes and shifts in pedestals 
caused by large signals in neighbouring cells were monitored and 
corrected using cells without signal in beam events~\cite{HardPaper}. The uncertainty on the pedestal levels was estimated 
to be less than 0.002~MIPs, negligible compared to the energy deposited by 
electron showers (a 10~GeV electron deposits on average 1450 MIPs).
The noise level was typically  0.13~MIPs; its spread channel-to-channel 
was  9\% of the mean noise  and the  spread run-to-run 
was less than 1\% of the mean noise. The low spread of the noise justifies the use of a single
energy  threshold for all cells in the detector.        
  
Calibration constants for each pad were determined using muon events.  
The response of each cell was fitted by a convolution of 
a Landau distribution with a Gaussian. The most probable value of the
underlying Landau function was taken to define the MIP value for each cell, 
and the raw energy for each cell in data was corrected to units of MIPs. 
All but 9 pads were functioning and successfully calibrated. The calibration
constants were determined with an accuracy of 0.5\% and had a cell to cell dispersion of 5\%. 
Data taken in the various beam test periods during summer and autumn 2006 
showed well correlated calibration constants,
with differences less than 1.6\%.   

One feature of the data which has not been accounted for in the detector simulation is associated with 
showers which deposit a sizeable energy in the guard ring surrounding a wafer.  
This is a cause of correlated crosstalk, observed as a 
distinctive square pattern of low energy hits in a number of cells 
around the periphery of the wafer.  The prevalence of this effect increases 
 with the shower energy crossing the guard ring and therefore its rate is significantly reduced  when considering only electrons impinging on the centre of the wafers. In the future, the design of the 
guard rings will be modified in order to prevent this problem. 

After calibration, the ECAL data consist of hits in 
the cells of the calorimeter with  energies in units of MIPs.  In order to remove most of the
noise signals, a threshold cut of 0.6~MIP was imposed on each cell, almost five times the
mean noise level.


\section{Monte Carlo Simulation}
\label{sec:MC}
The test beam setup is simulated with Mokka~\cite{MC:Mokka}, a Geant4~\cite{MC:G4}-based Monte Carlo program, followed by a digitisation module simulating the response of the data acquisition electronics. The material upstream of the ECAL is taken into account. The subdetectors are simulated with different levels of detail, depending on their impact on the physics analysis: material simulation only for the \v{C}erenkov detectors, raw energy depositions stored for
       the trigger counters, partial electronics simulation for the tracking detectors. In the case of the ECAL, the simulation gives the raw energy depositions in the Si pads and the readout electronics is simulated assuming that each channel exhibits only  Gaussian noise. The mean values of the noise for each channel follow a Gaussian distribution over the ECAL, with the mean value of 0.13~MIPs and 0.012~MIPs dispersion, as measured. Since 99.9\% of the ECAL cells were functioning, the impact of the non-responding cells is expected to be small and their signals were not supressed in the simulation  for this analysis.

The beam simulation assumes a parallel beam with Gaussian width reproducing the observed beam profile. 
To study systematic effects due to lateral leakage of the showers, 
samples are also generated with a beam spread uniformily
over the ECAL front face. A Gaussian momentum dispersion consistent with 
the settings of the beam collimators~\cite{TB:CERN} is 
applied for each run.


\section{Selection of Electron Events}
\label{sec:Selection}

\begin{figure}[hbtp]
\centering
  \epsfig{file=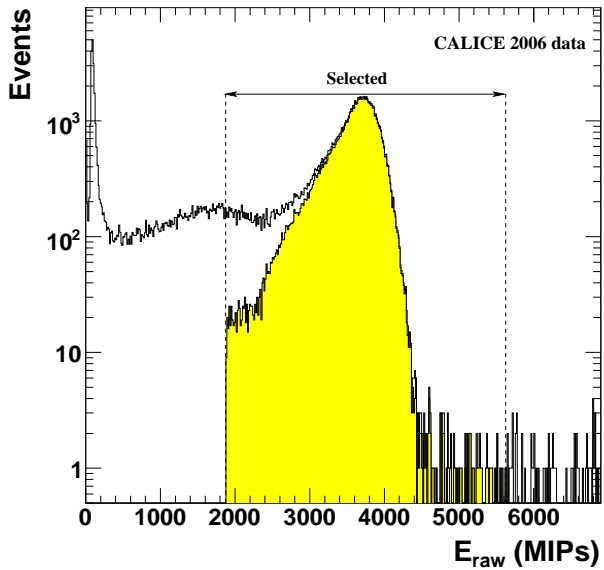, width=0.6\linewidth}
  \caption{ Distribution of total ECAL hit energies for a 15~GeV  electron run with a significant pion content. The $E_{\mathrm{raw}}$ selection window and
the shaded area obtained by demanding a
signal from the \v{C}erenkov counter are shown.}
\label{fig:Eraw}
\end{figure}

Single electron showers are selected using the energy 
recorded in the ECAL. This energy, $ E_{\mathrm{raw}}$, is 
calculated  with the three ECAL modules weighted in proportion to the 
tungsten thickness:
\begin{equation}
 E_{\mathrm{raw}}= \sum_{i=0}^{i=9}E_i + 2\sum_{i=10}^{i=19}E_i +3\sum_{i=20}^{i=29}E_i, 
\label{eq:Eraw}
\end{equation}
where $E_i$ is the energy deposit in layer $i$. The distribution of $E_{\mathrm{raw}}$ is shown in Figure~\ref{fig:Eraw} for a typical 15~GeV event sample. The electron peak at around  
3900~MIPs is clearly visible; however, the   muon and  pion contamination in the beam gives an 
additional peak at 85~MIPs and the region between the two main peaks is populated with pions.
Electron candidates are selected by requiring:
\begin{equation}
125 < \frac{E_{\mathrm{raw}}(\mathrm{MIP})}{E_{\mathrm{beam}}(\mathrm{GeV})}\;<  375.
\end{equation}
The significant pion contamination present in some of the data runs is reduced by demanding a trigger signal from the threshold \v{C}erenkov counter in the beam.  The effect of this additional requirement is indicated by the shaded region in Figure~\ref{fig:Eraw}. 

\subsection{Rejection of the beam halo}

The rejection of the beam halo is implemented run-by-run. 
The $x$ and $y$ acceptance for the incoming electron track 
is chosen such as to achieve a reasonably 
flat distribution of the mean energy deposition in the ECAL.

\subsection{Inter-wafer gap effect}
\begin{figure}[htbp]
  \centering
  \mbox{\epsfig{file=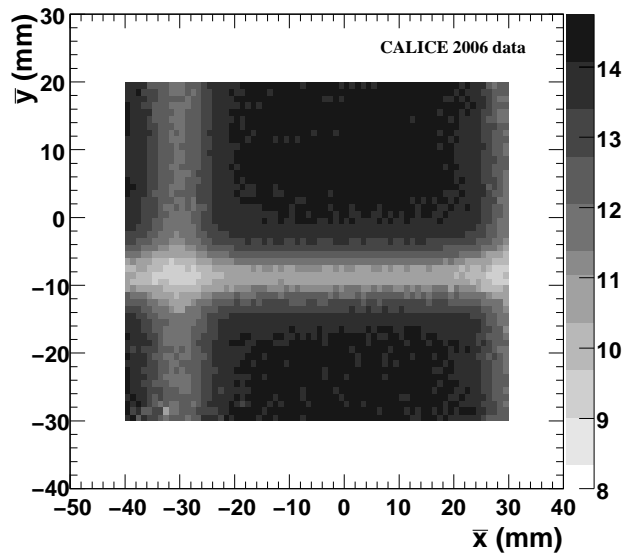, width=0.6\linewidth}}
 \caption{Mean values of $E_{\mathrm{raw}}$  for  15~GeV electrons as a function shower barycentre, transverse to the beam direction. The energies have been scaled down by a factor 266 to provide approximate conversion to GeV.}
\label{fig:gap}
\end{figure}

Around the pads in each wafer, a non-active region of 1~mm width was used for a grounded guard 
ring structure. This creates a non-active gap between adjacent Si pads situated on different wafers (2~mm) which is significant compared to the transverse shower size. These non-active regions, called in the following ``inter-wafer gaps'' degrade the prototype response when showers traverse them.
This is illustrated in Figure~\ref{fig:gap} for 30~GeV electrons impinging 
on the calorimeter at normal incidence. Here the mean value of $E_{\mathrm{raw}}$ (Equation~\ref{eq:Eraw}) is plotted 
as a function of the shower barycentre $(\bar{x}, \bar{y})$, defined as :
\begin{equation}
(\bar{x},\bar{y})=\sum_i(E_ix_i,E_iy_i)/\sum_iE_i 
\end{equation}
The sums run over all hit cells in the calorimeter. 
Dips in response corresponding to the guard ring positions are clearly visible: the energy loss is 
about 15\% when electrons impinge in the centre of the 
$x$ gaps and about 20\% in the case of the $y$ gap.
In order to recover this loss and to have a more uniform calorimeter 
response, a  simple method was investigated. The ECAL energy response, 
$f(\bar{x},\bar{y}) = E_{\mathrm{raw}}/E_{\mathrm{beam}}$,
is measured using a combined sample of 10, 15 and 20~GeV electrons, 
equally populated.

\begin{figure}[htbp]
  \centering
  \mbox{
    \epsfig{file=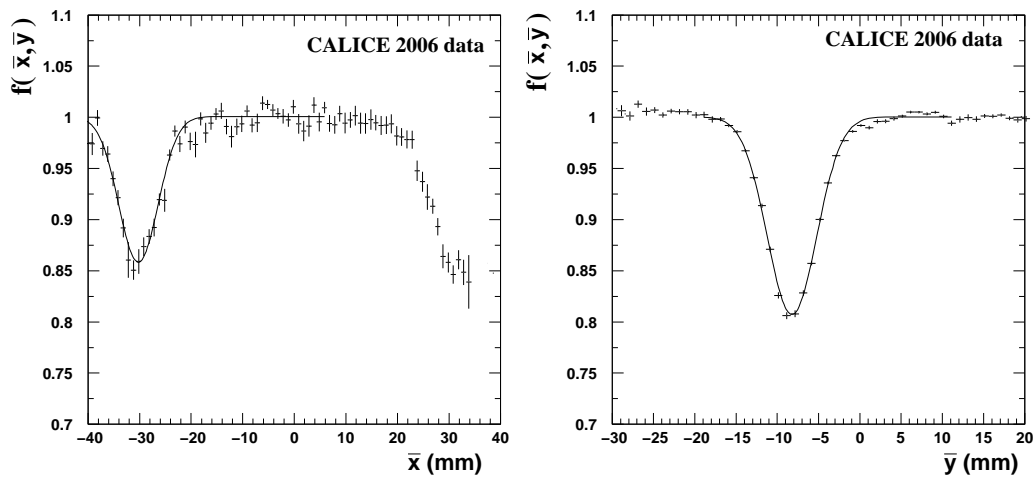,width=\linewidth}
  }
  \caption{Normalised $f(\bar{x},\bar{y})$ as a function of the shower barycentre coordinates, for a combined sample of 10, 15 and 20~GeV electrons.}
 \label{fig:global_gaps_correction_pardef} 
\end{figure}

The response function $f$, normalised such as to have a unit response in the middle of the wafers, is displayed on Figure~\ref{fig:global_gaps_correction_pardef}.  To characterise the $x$ ($y$) response, the events were required to be outside the inter-wafer gap in $y$ ($x$), leading to an important difference in the number of events for the two distributions, since the beam is  centred on the $y$ gap. It can be parametrised with Gaussian functions, independently in $\bar{x}$ and $\bar{y}$: 
\begin{equation}
  \label{eq:GaussianGaps}
  f(\bar{x},\bar{y}) = \left(1-a_{x} \mathrm{exp}\left(-\frac{ (\bar{x}-x_{{\mathrm{gap}}})^2 }{2\sigma_x^2}\right)\right)
\left(1-a_{y} \mathrm{exp}\left(-\frac{(\bar{y}-y_{{\mathrm{gap}}})^2}{2{\sigma}_{y}^2}\right)\right)
\end{equation}

Here, $x_{\mathrm{gap}}$ and $y_{\mathrm{gap}}$ are the positions at the centres of the gaps in $x$ and $y$, respectively, $a_x$ ($a_y$) and 
$\sigma_x$ ($\sigma_y$) their respective depths and widths in the two directions. The results of the Gaussian parametrisations  are given in 
Table~\ref{tab:global_gaps_correction_fits}.  The gap in $x$ is shallower 
and wider than that in $y$, due  to the staggering of the gaps in 
$x$~\cite{HardPaper}.

\begin{table}[htbp]
  \centering
  \begin{tabular}{|c|c|c|c|}
  \hline
                & position (mm) &  $\sigma$ (mm) &  \mbox{  }~$a$~\mbox{  } \\
  \hline
  $x$ direction & $-$30.0        &    4.3         & 0.143      \\
  \hline
  $y$ direction &   $-$8.4        &    \mbox{ }~3.2~\mbox{ }      & \mbox{ }~0.198~\mbox{ }      \\
  \hline
  \end{tabular}
  \caption{Gaussian parametrisation of the inter-wafer gaps.}
  \label{tab:global_gaps_correction_fits}
\end{table}

\begin{figure}[htbp]
  \centering
  \mbox{
    \epsfig{file=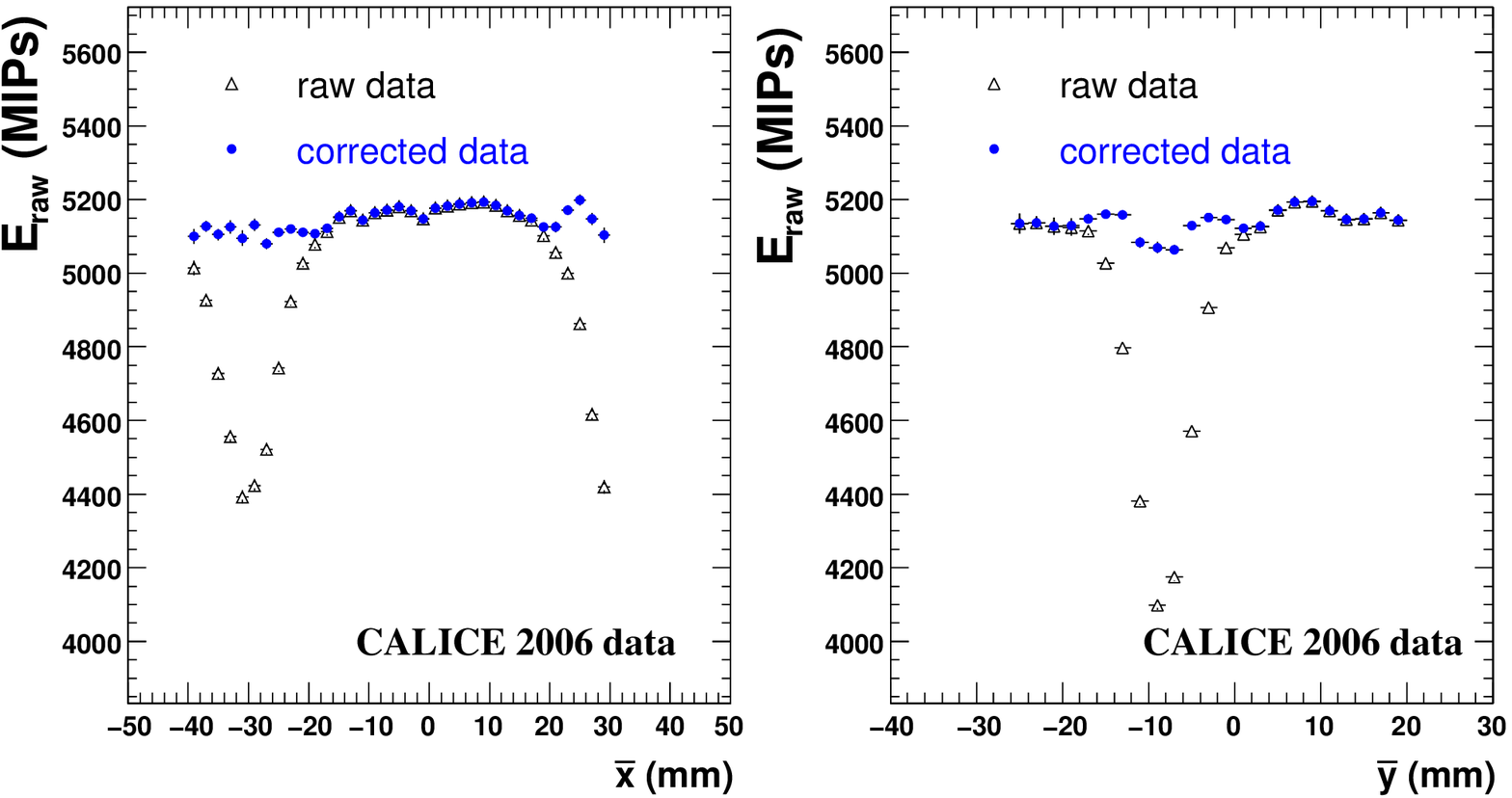,width=\linewidth}
  } 
  \caption{Mean $E_{\mathrm{raw}}$ as s function of the shower barycentre coordinates 
  for 20~GeV electrons, before  (open triangles) and after the inter-wafer gap corrections (solid circles) were applied on  $E_{\mathrm{raw}}$. }
 \label{fig:global_gaps_scan}
\end{figure}

\begin{figure}[htbp]
  \centering
  \mbox{
    \epsfig{file=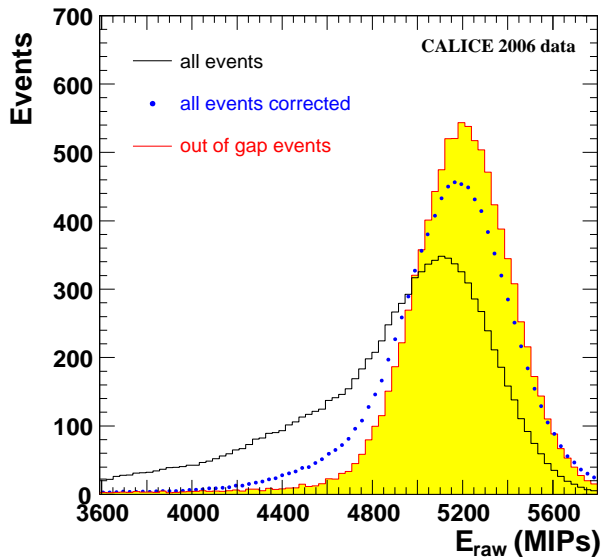,width=0.6\linewidth}
  } 
  \caption{Energy distribution for 20~GeV electrons in the cases of: events outside the inter-wafer gaps (solid histogram), all events without inter-wafer gap corrections (open histogram) and all events with inter-wafer gap corrections (solid circles). The histograms are normalised to the same number of entries. }
 \label{fig:global_gaps_energy}
\end{figure}

As illustrated in Figure~\ref{fig:global_gaps_scan}, when  the energy of each shower is corrected by $1/f$, the average energy loss in the gaps is reduced to a few 
percent level. The low energy tail  in the energy distribution is also  much reduced (Figure~\ref{fig:global_gaps_energy}). The correction method relies only on calorimetric information and can be applied both for photons and electrons. 

Even though it is possible to correct for the inter-wafer gaps on average, for individual events their presence will induce fluctuations in the energy response and degrade the ECAL resolution compared to a continuous calorimeter. 
In the data described here, the beam centre was close to the inter-wafer gap in $y$, artificially increasing the impact of the inter-wafer gaps compared to an experiment with a beam uniformely spread over the ECAL front face. Moreover, since the beam width varies strongly with the beam energy, the impact of the energy lost in the gaps is different at each energy. Therefore, in order to assess the energy response and resolution  of the prototype as a function of energy in a unbiased way, only particles  impinging in the middle of the wafers 
are  selected.  Since the gap effect 
is  negligible more than 4~standard deviations away from the gap 
centre,  the shower barycentre for selected events is 
required to be at a distance larger than 17.2~mm from the centre of the  inter-wafer gap  along $x$ and 12.76~mm away from the centre of the $y$ gap.

A sustained R\&D effort is being made to reduce the non-active areas, both by reducing the size of the inter-wafer gaps and   by increasing the size of the wafers. The next Si-W~ECAL prototype will have 9$\times$9 Si~pads in a wafer which leads to a significant  decrease of the  non-active areas.

\subsection{Selection of showers well contained in the ECAL}
\begin{figure}[hbtp]
\centering
  \mbox{
    \epsfig{file=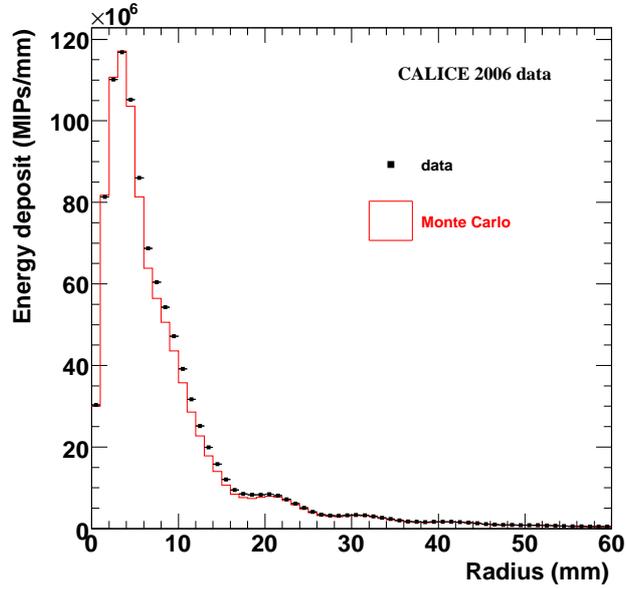, width=0.6\linewidth}
  }
  \caption{Energy deposited in ECAL as a function of the radial distance to the longitudinal shower axis, integrated over 171671 showers of 45~GeV.}
\label{fig:containment}
\end{figure}

The fiducial volume in which the showers are fully contained in the ECAL was estimated using 
electrons away from the inter-wafer gaps and pointing at the centre of the ECAL. The radial shape of an average 45~GeV electron shower is 
shown in Figure~\ref{fig:containment}, both for data and simulation. The simulation reproduces the shower width to better than 2\%: 
95\% of the shower energy is contained within 30.5~mm ({\it i.e.\ }less than four pads), to be compared with 29.9~mm in the case of the simulated showers.
 To ensure radial containment, all electrons impinging on the ECAL front face
less than 32~mm from one of the ECAL borders are therefore excluded from the selected sample.

The longitudinal containment of the showers is ensured by rejecting events which have the maximum of the energy deposited along the $z$ direction in the first five layers or the last five layers of the prototype. Only 0.21\%  of the simulated 6~GeV electrons fail these containment criteria and 0.02\% of the 45~GeV electrons.

\subsection{Rejection of electrons showering in front of ECAL}

\begin{figure}[hbtp]
\centering
  \epsfig{file=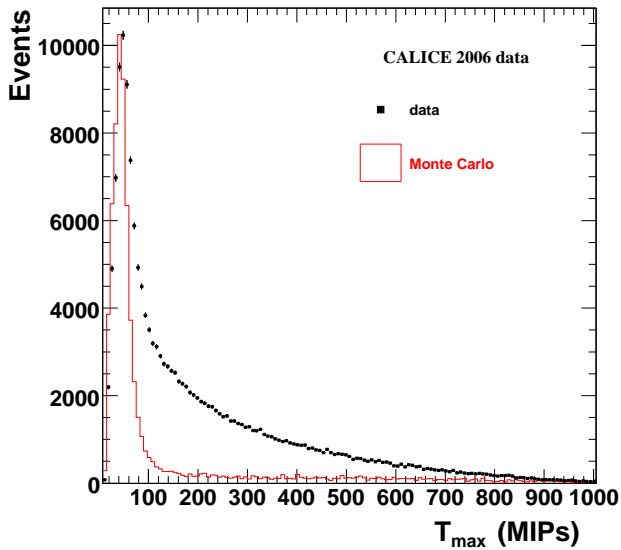, width=0.6\linewidth}
  \caption{Distribution of the variable $T_{\mathrm{max}}$ (described in the text), 
which is used to reduce the contribution of double showers.  Data and 
simulation are compared at 30~GeV.}
\label{fig:Tmax}
\end{figure}

The data recorded at CERN contain a significant number of events 
which have approximately the expected energy for a single electron, 
but whose spatial structure clearly exhibits double showers.  
A likely explanation is bremsstrahlung far upstream in the beamline.
In the Monte Carlo simulation, the known material between the 
\v{C}erenkov counter and the calorimeter is simulated, and yet the agreement 
between the rate of double shower events is poor between simulations and data.
Before comparing data and Monte Carlo, it is therefore necessary to select a sample of 
single electron showers.
To this end,  the energy deposits in the shower are projected in a two-dimensional histogram,  
on the transverse,  $x-y$, plane. The binning of the histogram is the same in $x$ and $y$ and corresponds to the cell size (1~cm). A simple  nearest-neighbour clustering algorithm  (including diagonal neighbours)  is applied on the bins with energies above a given  threshold $T$, in order to select events with more than one local maximum for the energy deposit. 
For each event we determine the maximum value of the threshold, 
$T_{\mathrm{max}}$, above which the event would be reconstructed as 
a single cluster.  In Figure~\ref{fig:Tmax} we compare the distributions 
of $T_{\mathrm{max}}$ between data and simulation for a 30~GeV beam.
A sizeable discrepancy is seen for larger values of $T_{\mathrm{max}}$
and therefore a cut is applied on $T_{\mathrm{max}}$.  The cut is energy 
dependent, varying from 50~MIPs at 10~GeV to 120~MIPs at 45~GeV.  This cut 
typically rejects $\sim$20\% of data and $\sim$2-3\% of simulated events 
at the higher energies.

A summary of the selected electron and positron data  is shown in Table~\ref{tab:tb_sum}. The number of simulated events available for each energy is also indicated.

\begin{table}[hbt]
\begin{center}
\begin{tabular}{|c|c|c|c| } \hline
Energy (GeV) & date      &  data statistics (kevts) & MC statistics (kevts)\\ \hline
6            & Oct       &  6.6                     & 83.2 \\ \hline
10           & Aug, Oct  &  43.1                    & 80.3 \\ \hline
12           & Oct       &  27.2                    & 72.8 \\ \hline
15           & Aug, Oct  &  51.4                    & 70.3 \\ \hline
20           & Aug       &  62.9                    & 56.2 \\ \hline
30           & Aug       &  42.3                    & 55.2 \\ \hline
40           & Aug       &  22.9                    & 67.8 \\ \hline
45           & Aug       &  108.6                   & 108.8\\ \hline
\end{tabular}
\end{center}
\caption{\sl Summary of the electron events selected for this analysis.}
\label{tab:tb_sum}
\end{table}


\section{Performance Studies}
\label{sec:Response}

\subsection{ECAL Sampling Fraction Scheme}
\begin{figure}[hbtp]
\centering
 \mbox{
    \epsfig{file=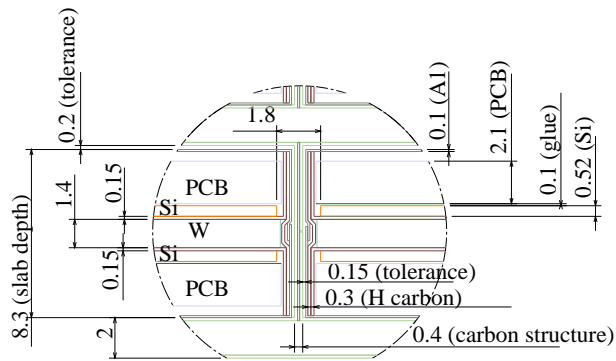, width=0.6\linewidth}
  }
  \caption{Details of one ECAL slab, showing one passive tungsten layer sandwiched between two active silicon layers. The dimensions are in mm.
               In contrast to the upper silicon layer which is preceded by a layer of tungsten only, the lower silicon layer is
               preceded by a larger passive layer: the PCB, aluminium, glue and carbon structure as well as the tungsten.}
\label{fig:protophys}
\end{figure}

The ECAL is made of 30 layers grouped in three modules of 10 layers each~\cite{HardPaper}. Each tungsten sheet has the same 
thickness in a given module. However, as can be derived from Figure~\ref{fig:protophys}, where one passive tungsten layer sandwiched between two active silicon layers is shown, two successive silicon layers are either separated by  one thickness of tungsten or by the same thickness of tungsten plus two thicknesses of PCB, aluminium  and
carbon-fibre--epoxy composite. A different sampling fraction, defined as the ratio of the energy deposited in the active medium to the total energy deposit (sum of the energy deposits in the active {\it and} passive medium), is therefore expected for the even and the odd layers  of the same calorimeter module.

The easiest method to investigate this difference is to compare in each module the mean energy deposits in odd and even layers. 
For the first module, if we neglect the shower profile, the ratio of the two is 
\begin{equation}
  R = \frac{E^{\rm odd}}{E^{\rm even}} = 1+\eta , 
\end{equation}
with $\eta$ being, approximately, the ratio of the non-tungsten radiation length to the tungsten radiation length.

\begin{figure}[hbtp]
\centering
  \mbox{
     \epsfig{file=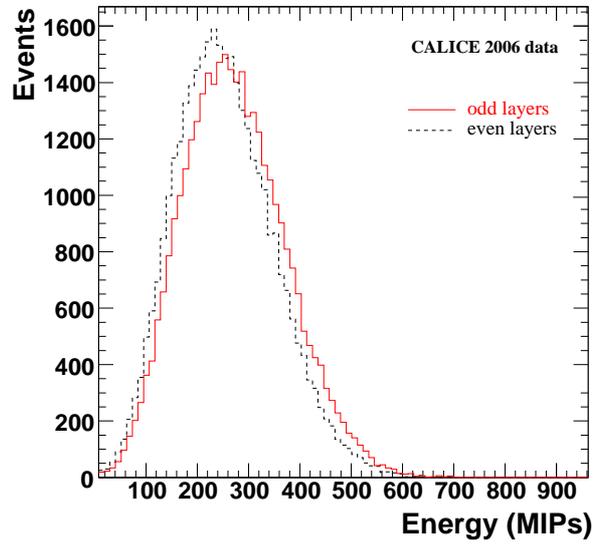, width=0.6\linewidth}
   }
  \caption{Energy deposits in odd layers (continuous histogram) and average energy deposits in their surrounding even layers (dashes) by 20~GeV electrons, in the first ECAL module.}
\label{fig:asym}
\end{figure}

When counting the layers starting from zero, the odd layers are systematically shifted compared to the even layers towards the shower maximum and the measurement of $R$ is biased by the shower development. To overcome this bias, $R$ is measured twice, either comparing
the odd layers with the average of the surrounding even layers, or
comparing the even layers with the average of the neighbouring odd
layers:
\begin{eqnarray}
R^\prime &=& \frac{ \left\langle{\rm E}_1+{\rm E}_3+{\rm E}_5+{\rm E}_7\right\rangle } { \left\langle\frac{{\rm E}_0+{\rm E}_2}{2}+\frac{{\rm E}_2+{\rm E}_4}{2}+ \frac{{\rm E}_4+{\rm E}_6}{2}+ \frac{{\rm E}_6+{\rm E}_8}{2} \right\rangle} \\
R^{\prime\prime}&=&\frac{ \left\langle\frac{{\rm E}_1+{\rm E}_3}{2}+\frac{{\rm E}_3+{\rm E}_5}{2}+ \frac{{\rm E}_5+{\rm E}_7}{2}+ \frac{{\rm E}_7+{\rm E}_9}{2}\right\rangle}{ \left\langle{\rm E}_2+{\rm E}_4+{\rm E}_6+{\rm E}_8\right\rangle}
\end{eqnarray}
where ${\rm E}_n$ is the energy deposit in the layer number $n$ and the brackets indicate that mean values are used.
The value of $\eta$ is taken as the average of $R^\prime-1$ and 
$R^{\prime\prime}-1$, 
while the difference between them gives a conservative estimate of the
 systematic uncertainty due to the shower shape.
As an example, the distributions of the energy deposits in the odd and even layers are shown in 
Figure~\ref{fig:asym} for $20$~GeV electrons.

\begin{figure}[hbtp]
\centering
  \mbox{
     \epsfig{file=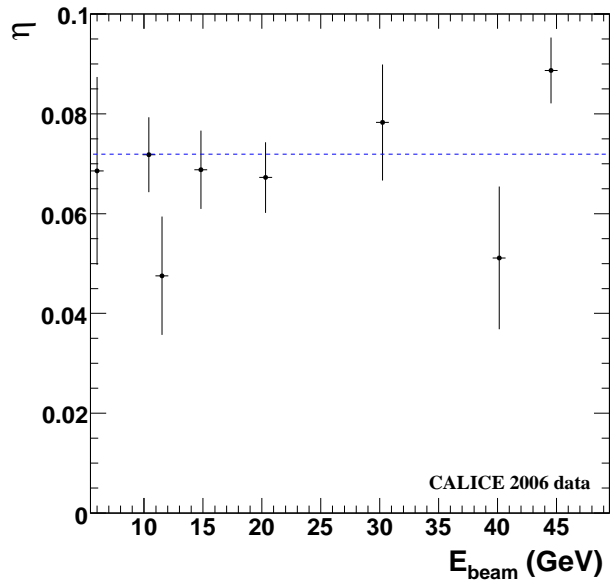, width=0.6\linewidth}
   }
  \caption{Values of $\eta$ as a function of the beam energy. The uncertainties are statistical and the dashed line gives the average value of $\eta$.}
\label{fig:eta}
\end{figure}

The values of $\eta$, obtained using the first module and for different beam energies, are  displayed in 
Figure~\ref{fig:eta}. The overall value is $(7.2\pm0.2\pm1.7)$\%.  The measurement of 
$\eta$ using the second and third module gives compatible results and the corresponding value obtained from simulation is $(4.7\pm0.2\pm2.0)$\%.

In computing the total response of the calorimeter, the sampling fraction
for layer $i$ is given by $w_i=K$  for even layers and $w_i=K+\eta$ for the odd layers, with
 $K=1,\ 2,\ 3$ in modules 1, 2, 3, respectively.

\subsection{Linearity and energy resolution}
\begin{figure}[htbp!]
  \centering
  \mbox{
    \epsfig{file=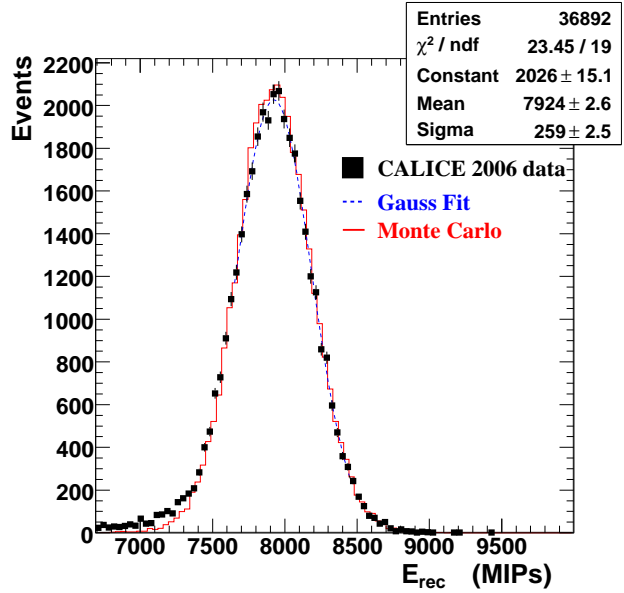, width=0.6\linewidth}
  }
  \caption{Gaussian parametrisation of $E_{\mathrm{rec}}$ for 30~GeV electron data (dashes). The range of the fit is $[-\sigma,+2\sigma]$. The data (solid squares) and simulation (continuous line) are superposed.}
\label{fig:Fit30}
\end{figure}

\begin{figure}[hbtp]
\centering
  \mbox{
     \epsfig{file=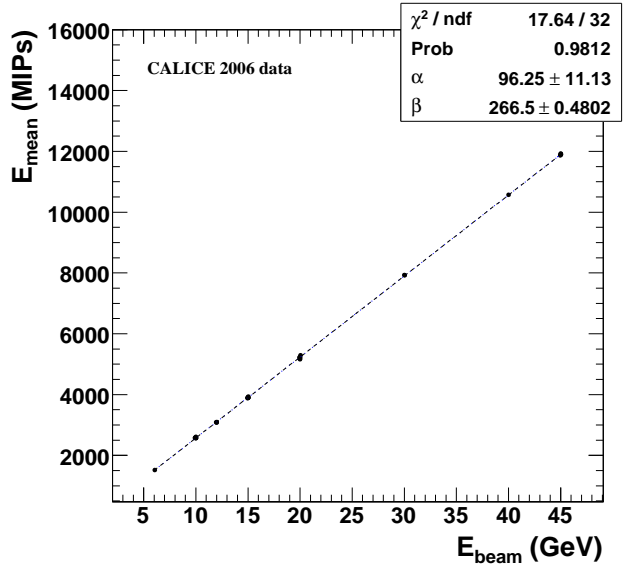, width=0.6\linewidth}
   }
   \caption{Energy response of the ECAL as a function of the beam energy. For clarity, all the runs around the same 
     nominal energy of the beam were combined in one entry for the plot, for which entry the uncertainty was estimated assuming that 
     the uncertainties on the individual runs were uncorrelated. }
\label{fig:linearity}
\end{figure}

The total response of the calorimeter is calculated as
\begin{equation}
 E_{\mathrm{rec}}(\mbox{MIPs})=\sum_i w_i E_i  
\end{equation}
with $w_i$ the sampling fraction for the layer $i$.
 Its distribution for electrons at 30~GeV is shown in Figure~\ref{fig:Fit30}, together with a fit
using a Gaussian function in the range $[-\sigma,+2\sigma]$. There is reasonably  good agreement between data and simulation.
An asymmetric range is chosen for the fit in order to reduce sensitivity to pion background, 
to radiative effects upstream of the calorimeter, and to any 
residual influence of the inter-wafer gaps.  The position of the peak 
is the mean energy response (called in the following $E_{\mathrm{mean}}$) and its distribution is shown in Figure~\ref{fig:linearity} as a function of the beam energy. The uncertainties on $E_{\mathrm{mean}}$ are those estimated from  the fit.

From the dispersion of $E_{\mathrm{mean}}$ in the different runs at the same nominal beam energy, the uncertainty of the beam mean energy, $E_{\mathrm{beam}}$, was estimated to be
\begin{equation}
  \frac{\Delta E_{\mathrm{beam}}}{E_{\mathrm{beam}}} = \frac{0.12}{E_{\mathrm{beam}}(\mathrm{GeV})}\oplus0.1\%,
\label{eq:beamerror}
\end{equation}
The first term is related to hysteresis in the bending magnets, while the calibration and the uncertainties on the collimator geometry give the constant term. For comparison, in~\cite{beamerror}, the uncertainty on the beam mean energy was quoted as
\begin{equation}
  \frac{\Delta E_{\mathrm{beam}}}{E_{\mathrm{beam}}} = \frac{0.25}{E_{\mathrm{beam}}(\mathrm{GeV})}\oplus0.5\%,
\label{eq:atlas}
\end{equation}
The first of these parametrisations of the uncertainty (Equation~\ref{eq:beamerror}) is used in the following, except for checks of  systematic uncertainties. 

\begin{figure}[hbtp]
\centering
  \mbox{
     \epsfig{file=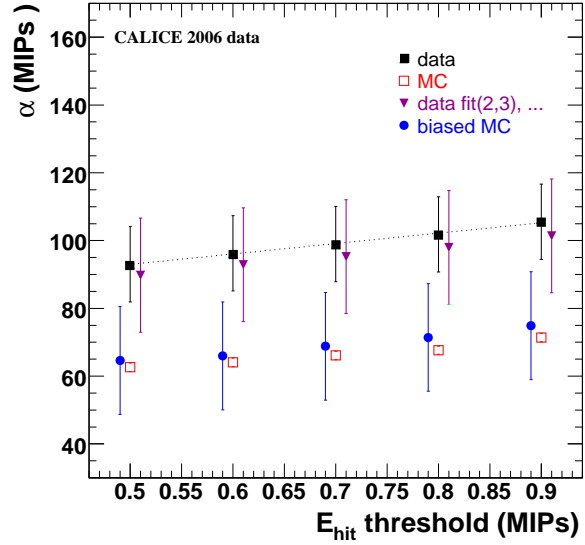, width=0.6\linewidth}
   }
   \caption{Variation of the linearity offset with the hit energy threshold: data (solid squares), data with the fitting range of $E_{\mathrm{rec}}$ enlarged to $[-2\sigma,+3\sigma]$ and considering the parametrisation in Equation~\ref{eq:atlas} for the uncertainty on the beam mean energy (solid triangles), Monte Carlo (open squares), Monte Carlo with an artificial uncertainty on the beam mean energy as given by Equation~\ref{eq:atlas} (solid circles). For clarity, the points were artificially shifted along the $x$ axis around the nominal $E_{\mathrm{hit}}$ threshold.}
\label{fig:linoffset}
\end{figure}

The  mean energy response can be parametrised as 
$E_{\mathrm{mean}} = \beta\cdot E_{\mathrm{beam}} - \alpha$, while the $measured$ energy $E_{\mathrm{meas}}$ is given by
$E_{\mathrm{meas}}= E_{\mathrm{mean}} +\alpha$. The parameter
$\beta$ is a global MIP to GeV calibration factor. The offset $\alpha$ is partly due to the rejection of the low energy hits and  it increases steadily  with the hit energy threshold, as  displayed on Figure~\ref{fig:linoffset}. On the same figure are also shown the values of the offset, as expected from simulation. The bias introduced by the fact that the uncertainty on the beam mean energy decreases with increasing beam energy and  enhances therefore the weight of the high energy runs was estimated by artificially assigning to the simulated data uncertainties on $E_{\mathrm{beam}}$ according to~Equation~\ref{eq:beamerror}. By taking into account this bias, the disagreement between data and simulation is somewhat reduced (Figure~\ref{fig:linoffset}).

\begin{figure}[hbtp]
\centering
  \mbox{
     \epsfig{file=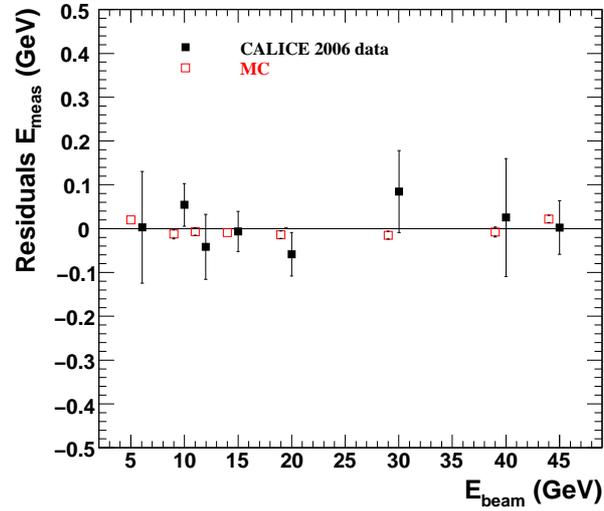, width=0.6\linewidth}
   }
   \caption{Residuals to linearity of $E_{\mathrm{meas}}$ as a function of  the beam energy, for data (solid squares) and simulation (open squares). All the runs around the same nominal energy of the beam were combined in one entry, for which the uncertainty was estimated assuming that the uncertainties on the individual runs were uncorrelated. For clarity, the Monte Carlo points were artificially shifted along the $x$ axis around the nominal $E_{\mathrm{beam}}$.}
\label{fig:linresid}
\end{figure}

The residuals to linearity of the measured energy, converted from MIPs to GeV using a constant 266 MIPs/GeV conversion factor (obtained from Figure~\ref{fig:linearity}), are shown in Figure~\ref{fig:linresid} as a function of the beam energy. The residuals  are   within approximately the 1\% level and are consistent with zero non-linearity. Data and simulation  agree within one standard deviation.

\begin{figure}[hbtp]
\centering
  \mbox{
     \epsfig{file=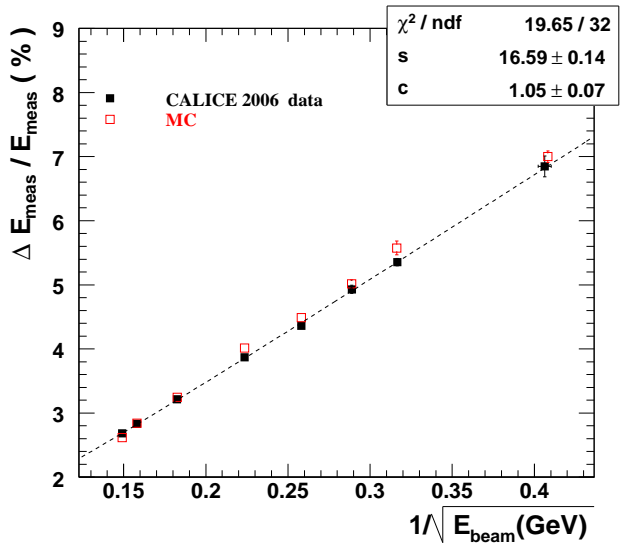, width=0.6\linewidth}
   }
   \caption{Relative energy resolution ($\Delta E_{\mathrm{meas}}/E_{\mathrm{meas}}$) as a function of the beam energy (solid squares), and its usual parametrisation as $s/\sqrt{E} \oplus c$. For clarity, the  35 runs available were combined into 8 different beam energy points for the plot. For the parametrisation of the energy resolution each run was however treated individually. The values expected from simulation are superposed (open squares).}
\label{fig:Resolution}
\end{figure}

The relative energy resolution, $\Delta 
E_{\mathrm{meas}}/E_{\mathrm{meas}}$ , 
as  shown in Figure~\ref{fig:Resolution},  can be 
parametrised by a quadrature sum of stochastic and constant terms 
\begin{equation}
 \frac{\Delta E_{\mathrm{meas}}}{E_{\mathrm{meas}}}= 
\left( \frac{16.6\pm0.1}{\sqrt{E(\mathrm{GeV})}} 
\oplus \left(1.1\pm0.1\right)\right)\%,  
\end{equation}
where  the intrinsic momentum spread of the beam was subtracted from the ECAL data~\cite{TB:CERN}. By varying the range for fitting $E_{\mathrm{rec}}$, the variation of the stochastic term has the same order of magnitude as the statistical error, whereas the constant term remains stable: a fitting interval reduced to $[-0.75\sigma,+1.75\sigma]$ improves the stochastic term to $16.5\pm0.2$, whereas an enlargement to $[-2\sigma,+3\sigma]$ degrades it to $16.7\pm0.1$. The expected resolution  from simulation is
\begin{equation}
\left[\frac{\Delta E_{\mathrm{meas}}}{E_{\mathrm{meas}}}\right]^{\mathrm{MC}} = \left(
\frac{17.3\pm0.1}{\sqrt{E(\mathrm{GeV})}} 
\oplus \left(0.5\pm0.1\right)\right) \%
\end{equation}
which agrees within  5\% with the measured resolution of the prototype. 

Different systematic checks have been performed on the data. Variations of the linearity and resolution against the minimal accepted distance between the shower 
barycentre and the nearest inter-wafer gap, when the  energy threshold  for considering the hits is  0.6~MIPs are shown 
in Table~\ref{tab:systematics1}.
 In addition, this hit energy threshold has itself been   varied (Table~\ref{tab:systematics2}).  In order to investigate the potential effects linked to the beam position, the energy response is also compared for showers with barycentres located in the  right  hand side (negative $x$ coordinates) and in the upper half of the detector (upper row of wafers) as summarised in Table~\ref{tab:systematics3}. The results of all checks are consistent. Since data were taken in both August and October 2006, it was also possible to check the response stability in time and no significant differences between the two data samples are observed.

\begin{table}[hbt]
\begin{center}
\begin{tabular}{|c|c|c|c|c|} \hline
                 & \multicolumn{4}{|c|}{shower distance to the gaps (in standard deviations)  }  \\   \cline{2-5} 
                 & 3.5               & 4                & 4.5             & 5                \\  \hline
 $\chi^2/ndf$    &  16.8/32          &  17.6/32         & 18.9/32         & 24.2/32          \\  
(linearity)      &                   &                  &                 &                  \\ \hline
 $\alpha$        &  93.9$\pm$11.1    &  96.3$\pm$11.2   &  97.8$\pm$11.5  & 99.1$\pm$11.6   \\ 
 (MIPs)          &                   &                  &                 &                  \\ \hline
 $\beta$         & 266.3$\pm$0.5     & 266.6$\pm$0.5    &  266.8$\pm$0.5  & 266.8$\pm$0.5    \\  
(MIPs/GeV)       &                   &                  &                 &                  \\ \hline
 resolution (\%) & 16.7$\pm$0.1      & 16.6$\pm$0.1     &  16.4$\pm$0.2   & 16.3$\pm$0.2     \\  
(stochastic term)&                   &                  &                 &                  \\ \hline
 resolution (\%) & 1.0$\pm$0.1       &  1.0$\pm$0.1     &  1.1$\pm$0.1    & 1.2$\pm$0.1      \\  
(constant  term) &                   &                  &                 &                  \\ \hline
\end{tabular}
\end{center}
\caption{Impact of the distance of shower to the inter-wafer gaps on the ECAL linearity and resolution. The distance is given in terms of standard deviations to the gap centre, with the standard deviation defined by the Gaussian parametrisation of the gaps.}
\label{tab:systematics1}
\end{table}

\begin{table}[hbt]
\begin{center}
\begin{tabular}{|c|c|c|c|} \hline
                   & \multicolumn{3}{|c|}{$E_{\mbox{hit}}$ cutoff (MIPs)}\\ \cline{2-4} 
                   &    0.5           & 0.7              & 0.9             \\  \hline
 $\chi^2/ndf$      & 18.0/32          & 17.8/32          & 18.0/32         \\
(linearity)        &                  &                  &                 \\ \hline
 $\alpha$          &  93.0 $\pm$11.2  &  98.9$\pm$11.1   &  105.6$\pm$11.1 \\ 
 (MIPs)            &                  &                  &                 \\ \hline
 $\beta$           & 266.8$\pm$0.5    & 266.3$\pm$0.5    &  265.8$\pm$0.5  \\  
(MIPs/GeV)         &                  &                  &                 \\ \hline
 resolution (\%)   & 16.6$\pm$0.1     & 16.5$\pm$0.1     & 16.6$\pm$0.1    \\  
(stochastic term)  &                  &                  &                 \\ \hline
 resolution (\%)   & 1.0$\pm$0.1      & 1.1$\pm$0.1      & 1.1$\pm$0.1      \\  
(constant term)    &                  &                  &                \\ \hline
\end{tabular}
\end{center}
\caption{Impact of the hit energy cutoff on the ECAL linearity and resolution. }
\label{tab:systematics2}
\end{table}

\begin{table}[hbt]
\begin{center}
\begin{tabular}{|c|c|c|} \hline
                   & right side       & upper part      \\  \hline
 $\alpha$ (MIPs)   &  96.1$\pm$10.9   &  97.7$\pm$11    \\ \hline
 $\beta$ (MIPs/GeV)&   266.6$\pm$0.5  &  266.8$\pm$0.5  \\ \hline
 resolution (stochastic term) (\%)&   16.8$\pm$0.1   &  16.8$\pm$0.2   \\ \hline    
 resolution  (constant term)  (\%)&   1.1$\pm$0.1    &  1.1$\pm$0.1    \\ \hline
\end{tabular}
\end{center}
\caption{Response to electrons crossing the right hand side and the upper part of the ECAL.}
\label{tab:systematics3}
\end{table}


\section{Conclusion}

The response to normally incident electrons of the CALICE Si-W electromagnetic calorimeter was  measured for energies between 6 and 45~GeV, using the data recorded in 2006  at CERN.

The calorimeter response is linear to within approximately 1\%. The energy resolution has a stochastic term of \mbox{$(16.6 \pm 0.1)\%/ \sqrt{E({\mathrm{GeV}})}$}, whereas the constant term is $1.1\pm 0.1$\%. Several sources of systematic uncertainties have been investigated and their effect is within the statistical uncertainties. The agreement between data and Monte Carlo simulation is within 5\%.


\section*{Acknowledgements}

We would like to thank the technicians and the engineers who
contributed to the design and construction of the prototypes, including 
U.~Cornett, G.~Falley, K~Gadow, 
P.~G\"{o}ttlicher, S.~Karstensen and P.~Smirnov. We also
gratefully acknowledge the DESY and CERN managements for their support and
hospitality, and their accelerator staff for the reliable and efficient
beam operation. 
We would like to thank the HEP group of the University of
Tsukuba for the loan of drift chambers for the DESY test beam.
The authors would like to thank the RIMST (Zelenograd) group for their
help and sensors manufacturing.
This work was supported by the 
Bundesministerium f\"{u}r Bildung und Forschung, Germany;
by the DFG cluster of excellence ``Origin and Structure of the Universe'';
by the Helmholtz-Nachwuchsgruppen grant VH-NG-206;
by the BMBF, grant numbers 05HS6VH1 and 05HS6GU1;
by the Alexander von Humboldt Foundation (Research Award IV, RUS1066839 GSA);
by joint Helmholtz Foundation and RFBR grant HRJRG-002, Russian Agency for Atomic Energy, ISTC grant 3090;
by Russian Grants  SS-1329.2008.2 and RFBR0402/17307a
and by the Russian Ministry of Education and Science;
by CRI(MST) of MOST/KOSEF in Korea;
by the US Department of Energy and the US National Science
Foundation;
by the Ministry of Education, Youth and Sports of the Czech Republic
under the projects AV0 Z3407391, AV0 Z10100502, LC527  and by the
Grant Agency of the Czech Republic under the project 202/05/0653;  
and by the Science and Technology Facilities Council, UK.







\begin{thebibliography}{00}





\bibitem{ILC-RDR} T. Behnke {\it et al} (ed.), Reference Design Report ``Volume~4: Detectors'', available at 

{\verb http://lcdev.kek.jp/RDR }



\bibitem{pflow} J.-C. Brient, ``Improving the jet reconstruction with the particle flow method: An
introduction'', in the Proceedings of 11th International Conference on Calorimetry in High-Energy Physics (Calor 2004),
Perugia, Italy, March 2004.

\bibitem{HardPaper} M. Anduze {\it et al}, ``Design and Commissioning of the Physics Prototype of a Si-W Electromagnetic Calorimeter for the International Linear Collider'', JINST 3 (2008) P08001, 33p.

\bibitem{TB:HCAL} F. Sefkow, ``The scintillator HCAL testbeam prototype'', in the Proceedings of 2005 International Linear Collider Workshop (LCWS 2005), Stanford, California, March 2005.

\bibitem{TB:TCMT} D. Chakraborty, ``The Tail-Catcher/Muon Tracker for the CALICE Test Beam''', in the Proceedings of 2005 International Linear Collider Workshop (LCWS 2005), Standford, California, March 2005. 






\bibitem{TB:CERN} Description of the CERN H6 testbeam area available at\\
{ \verb http://ab-div-atb-ea.web.cern.ch/ab-div-atb-ea/BeamsAndAreas/ }





\bibitem{MC:Mokka} ``Mokka Geant4 Application for Linear Collider Detectors'', see\\
{\verb http://polzope.in2p3.fr:8081/MOKKA. }
\bibitem{MC:G4} S. Agostinelli {\it et al}, ``Geant4 --- A Simulation Toolkit'', NIM A 506 (2003) 250-303.


\bibitem{beamerror} S. Akhmadaliev {\it et al}, ``Results from a new combined test of an electromagnetic liquid argon calorimeter with a hadronic scintillating-tile calorimeter'', NIM A 449 (2000) 461-477.





\end{thebibliography}
\end{document}